\documentclass[aps,pra,twocolumn,superscriptaddress,longbibliography]{revtex4-1}

\usepackage{float,amsmath,amssymb,bbm,mathrsfs,bm,braket,color,graphicx,comment,amsfonts,dsfont,mathrsfs}
\usepackage[colorlinks,citecolor=blue,urlcolor=blue]{hyperref}
\usepackage[mathscr]{euscript}
\usepackage[normalem]{ulem}
\usepackage{comment}
\usepackage{dsfont}
\usepackage{xfrac}

\DeclareMathAlphabet\mathbfcal{OMS}{cmsy}{b}{n}

\begin{document}

\title{Non-Hermitian skin effect of dislocations and its topological origin}

\author{Balaganchi A. Bhargava}
\affiliation{Institute for Theoretical Solid State Physics, IFW Dresden and W\"urzburg-Dresden Cluster of Excellence ct.qmat, Helmholtzstr. 20, 01069 Dresden, Germany}

\author{Ion Cosma Fulga}
\affiliation{Institute for Theoretical Solid State Physics, IFW Dresden and W\"urzburg-Dresden Cluster of Excellence ct.qmat, Helmholtzstr. 20, 01069 Dresden, Germany}

\author{Jeroen van den Brink}
\affiliation{Institute for Theoretical Solid State Physics, IFW Dresden and W\"urzburg-Dresden Cluster of Excellence ct.qmat, Helmholtzstr. 20, 01069 Dresden, Germany}
\affiliation{Institute  for  Theoretical  Physics,  TU  Dresden,  01069  Dresden,  Germany}

\author{Ali G. Moghaddam}
\affiliation{Department of Physics, Institute for Advanced Studies in Basic Sciences (IASBS), Zanjan 45137-66731, Iran}
\affiliation{Computational Physics Laboratory, Physics Unit, Faculty of Engineering and
Natural Sciences, Tampere University, P.O. Box 692, FI-33014 Tampere, Finland}
\affiliation{Institute for Theoretical Solid State Physics, IFW Dresden and W\"urzburg-Dresden Cluster of Excellence ct.qmat, Helmholtzstr. 20, 01069 Dresden, Germany}

\date{\today}

\begin{abstract}
We demonstrate that dislocations in two-dimensional non-Hermitian systems can give rise to density accumulation or depletion through the localization of an extensive number of states. 
These effects are shown by numerical simulations in a prototype lattice model and expose a different face of non-Hermitian skin effect, by disentangling it from the need for boundaries. 
We identify a topological invariant responsible for the dislocation skin effect, which takes the form of a ${\mathbb Z}_2$ Hopf index that depends on the Burgers vector characterizing the dislocations. 
Remarkably, we find that this effect and its corresponding signature for defects in Hermitian systems falls outside of the known topological classification based on bulk-defect correspondence. 
\end{abstract}
\maketitle

\emph{Introduction}.---  
The Hermiticity of observables in quantum mechanics is a concrete part of its mathematical structure and physical interpretations.
However, upon dealing with dissipative quantum systems as well as a variety of classical platforms such as electrical circuits, photonic crystals and mechanical metamaterials, non-Hermitian (NH) Hamiltonians can effectively describe an essential part of their dynamics \cite{Bender2007, Rotter2009, elganainy2018, Ozawa2019,Ueda2020review}.
Recently, in light of overwhelming interest and progress in understanding topological phases, NH models have been also revisited from a topological point of view \cite{Ueda2018, alvarez2018}.
Although they share certain similarities with Hermitian systems, NH systems possessing complex energy spectra can reveal quite distinct topological features \cite{bergholtz2019review, Sato2019}. 
In particular, it has been found that in absolute contrast to any Hermitian system, in NH models the presence of open boundaries can drastically affect the energy spectra and subsequently, all eigenstates accumulate towards one end of the system, a phenomenon called the non-Hermitian skin effect (NHSE) \cite{lee2016, Yao2018, FoaTorres2018, Thomale2019,Song2019nhse, Thomale2020generalized,Longhi2019, wang2019realspace,xiao2020non, Weidemann2020, Neupert2020-PRR, Hughes2020skin, Hughes2020skinHall,li2020critical,Longhi2020unraveling,Yang2020Chiral,Chen2020,Gong2020loss,ryu2020Fieldtheory}. 
This effect, which necessitated revisiting the concept of bulk-boundary correspondence in Hermitian topological phases \cite{Xiong2018, Bergholtz2018, Regnault2019, slager2020}, originates from nontrivial point gap topology, represented by the one-dimensional (1D) winding number of the complex energy bands \cite{Sato2020,Zhang2020Correspondence}.

Here, we address the question of how the NHSE manifests in the presence of topological defects such as dislocations, instead of open boundaries. 
Many studies have revealed the presence of topologically protected modes at defects in Hermitian topological systems \cite{ran2009, juricec2012, slager2014, teo2013, benalcazar2014, deJuan2014, li2018topological, ortix2018, queiroz2019, geier2020, peterson2020}. 
The topological defects have been thoroughly classified by Teo and Kane, who have put forward the concept of bulk-defect correspondence \cite{teo2010, teo2017, chiu2016}.
The classification of topological defects has been also extended to NH systems \cite{Chen2019classification}, and very recent works have examined defect modes in NH topological phases \cite{Moessner2021dislocation, hughes2021disclination, Schindler2021}. 
However, the interplay between nontrivial defects and point gap topology, the latter being an intrinsically NH feature, still remains to be understood.

\begin{figure}[t]
\includegraphics[width=0.99\columnwidth]{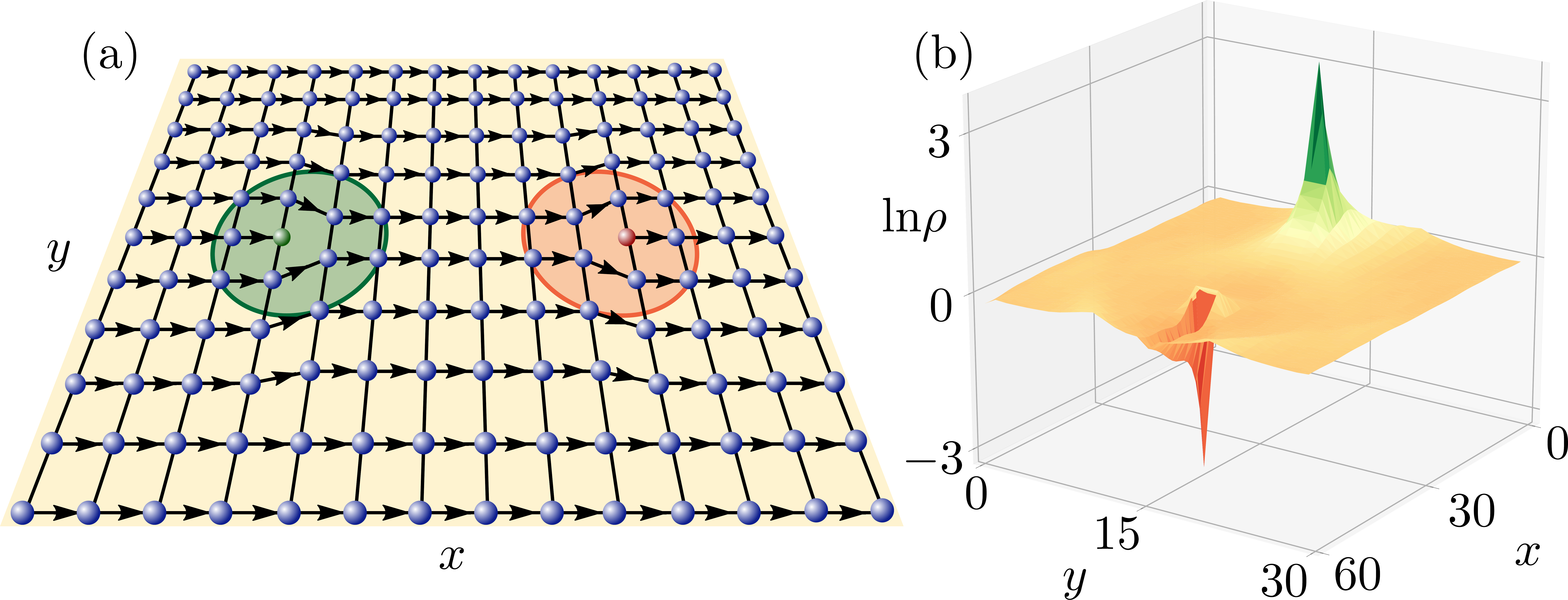} 
\caption{
(a) Sketch of the WHN model with two dislocations which are indicated by encircling them. 
(b) Logarithm of local density profile of the system with $L=30$ and under periodic boundary conditions and using a logarithmic scale for the density. 
The green peak and red dip represent the skin and anti-skin effects in which extensive number of exponentially localized states are piled up and depleted from the two dislocations, respectively. 
\label{fig:WHN}}
\end{figure}

By considering a prototype two-dimensional (2D) NH lattice model, we find both macroscopically large density accumulation and also density depletion in the vicinity of dislocations, which we respectively call skin and anti-skin effects (see Fig. \ref{fig:WHN}). 
This observation not only shows an unexplored consequence of the NHSE effect, but strikingly it is related to a Hermitian topological system hosting a new type of bulk-defect correspondence, beyond the known \emph{stable} classifications \cite{teo2010, teo2017}. 
Nevertheless, we identify a topological invariant $\vartheta=\hat{\bf z}\cdot({\bm \nu}\times {\bf B})/2$ to diagnose the macroscopic density collapse in terms of the Burgers vector ${\bf B}$ of the dislocation and two weak indices, ${\bm\nu}=({\nu_x,\nu_y})$, which are the 2D averaged winding numbers. 
We show that the topological invariant $\vartheta$ has an integer/half-integer parity given by the topological $\theta$ term.
{We further provide an intuitive picture for $\vartheta$ in connection with
topological invariants of weak topological phases of Hermitian systems subjected to topological defects.}

\emph{Weak Hatano-Nelson model}.--- 
As a preliminary, we first consider the simplest prototypical model which shows the NHSE, the Hatano-Nelson (HN) model described by the Hamiltonian ${\cal H}_{\rm 1dHN} = \sum_{j,\eta=\pm 1} (t + \eta \,\delta t)\, c^\dag_{j+\eta} c_j$ as a single-orbital tight-binding chain with nonreciprocal hoppings \cite{Hatano1996,Hatano1997}. 
A nonzero winding number of energy bands given by
\begin{equation}\label{eq:nuHN1d}
 \nu = \frac{1}{2\pi i} \int_{-\pi}^{\pi} d k \frac{\partial \varepsilon(k)}{\partial k} \frac{1}{\varepsilon(k)} = {\rm sign} (\delta t) = \pm1.
\end{equation}
signals the existence of the NHSE, revealed by the sensitivity of all eigenstates to the boundary conditions and the breakdown of the conventional bulk-boundary correspondence.
The NHSE can be characterized in terms of the localization of the probability density defined as $\rho_{\bf r}=\sum_{n}\big|\langle {\bf r}|\psi^R_{n} \rangle \big|^2$ in which $| \psi^R_n \rangle$ is the $n^{\rm th}$ right-eigenstate of the Hamiltonian, $| {\bf r} \rangle$ is the position ket, and the summation is over all eigenstates regardless of eigenvalue.
For a 1D HN model with open boundaries, all eigenstates are localized towards one end and as a consequence, the probability density is $\rho_r\propto e^{r/\xi_{\rm 1D}}$ with a 1D localization length $\xi_{\rm 1D}=1/\log \big[(t+\delta t)/(t-\delta t)\big]$ determined by the strength of non-reciprocity, $\delta t$.

We construct a 2D model by stacking HN chains along the $y$ direction, with nearest neighbor inter-chain hoppings as schematically shown in Fig.~\ref{fig:WHN}(a), and with the Hamiltonian
\begin{equation}\label{eq:HN2d}
{\cal H} = \sum_{\bf r}\sum_{\eta=\pm1} 
\big[(t_x + \eta \:\delta t_x)\: c^\dag_{{\bf r} + \eta\hat{\bf x}} c_{\bf r} 
+
t_y\: c^\dag_{{\bf r} + \eta\hat{\bf y}} c_{\bf r}\big],
\end{equation}
in which only hoppings in the $x$ direction are nonreciprocal \footnote{All numerical results are obtained using the Kwant code \cite{Groth2014}.}. 
We refer to this Hamiltonian as the \emph{weak} HN model (WHN), in analogy to weak topological insulators which are formed by stacking lower-dimensional strong topological insulators. 
Its eigenvalues form a family of ellipses in the complex plane which wind around the origin as a function of momentum $\varepsilon_{\bf k} = 2 \,\big( t_x\,\cos k_x + t_y\,\cos k_y+i \,\delta t_x \,\sin k_x\big)$.
In momentum space, the Hamiltonian ${\cal H}_{\bf k}$ corresponding to Eq.~\eqref{eq:HN2d} can be considered as a stack of decoupled 1D HN models with a $k_y$-dependent chemical potential $2t_y\cos k_y$.
We introduce weak indices
\begin{eqnarray}
\nu_j = \int \frac{d^2{\bf k}}{i(2\pi)^2}\: {\cal H}_{\bf k}^{-1} \,\partial^{\phantom\dag}_{k_j}{\cal H}_{\bf k}^{\phantom\dag} 
\label{eq:average-windings}, 
\end{eqnarray}
as the average of two 1D winding number densities $\nu(k_i)$ along the $j=x,y$ directions.
As long as $|t_y|<|t_x|$, there exists a point gap at $\varepsilon=0$ in the spectrum. 
Subsequently, the weak indices Eq.~\eqref{eq:average-windings} are quantized, and for our particular stacked model we obtain $\nu_x={\rm sign}(\delta t_x)$ and $\nu_y=0$. 
Consequently, for a system with open boundaries all states show a NHSE along the $x$ direction and, depending on the sign of $\delta t_x$, they localize either on the left or the right edge. 

\emph{Dislocation-induced skin effect}.--- 
We consider a rectangular system consisting of $2L\times L$ sites in the $x$ and $y$ directions, imposing periodic boundary conditions (PBC) in order to form a torus and introduce two dislocations separated by a length of $L$ sites using the cut and glue procedure.
A row of $L$ sites is first removed from the system, and the resulting cut is glued back together using vertical hoppings $t_y$, as shown in Fig.~\ref{fig:WHN}(a).

We find that the density localizes at one of the dislocations, forming a peak, and depletes at the other one, forming a dip [see Fig.~\ref{fig:WHN}(b), obtained for $t_y/t_x=0.4$, $\delta t_x/t_x=0.6$]. 
This behavior is indeed different from the conventional NHSE, although the underlying mechanisms are the non-reciprocity of the hopping parameters in both cases. 
First, the localized density depletion around the second dislocation (anti-skin effect) {is a feature of the NHSE which does not occur in uniform, defect-free NH systems.}
Second, in spite of the fact that vertical hopping terms are reciprocal, the exponential accumulation and depletion of the density around the dislocations take place in both the $x$ and the $y$ directions, as can be seen in Fig.~\ref{fig:WHN}(b).

The formation of skin and anti-skin effects at the dislocations can be understood intuitively in two different ways. 
The first is to consider the decoupled limit, $t_y=0$, in which the system consists of independent 1D Hatano-Nelson models. 
In this case, introducing dislocations amounts to removing half of the sites from one of the HN chains, such that it goes from periodic to open boundary conditions.
This leads to the formation of a NHSE at its boundaries [red and green sites in Fig.~\ref{fig:WHN}(a)].
However, the skin and anti-skin effects are present also when $t_y\neq 0$. 
This fact can be understood as arising from the nonreciprocal hoppings in the horizontal direction.
For $\delta t_x>0$, the hopping to the right is larger than the hopping to the left, implying that states will be pumped in the direction of the larger hopping \cite{Lee2019}, as indicated by arrows in Fig.~\ref{fig:WHN}(a).
Examining a closed loop surrounding each of the two dislocations, shown by green and red circles, we find that the number of arrows pointing into the contour is different from the number pointing out of the contour.
This imbalance is consistent with the density accumulation and depletion shown in Fig.~\ref{fig:WHN}(b), suggesting that states are pumped into or out of the dislocations cores.

\begin{figure}[t]
\includegraphics[width=0.99\columnwidth]{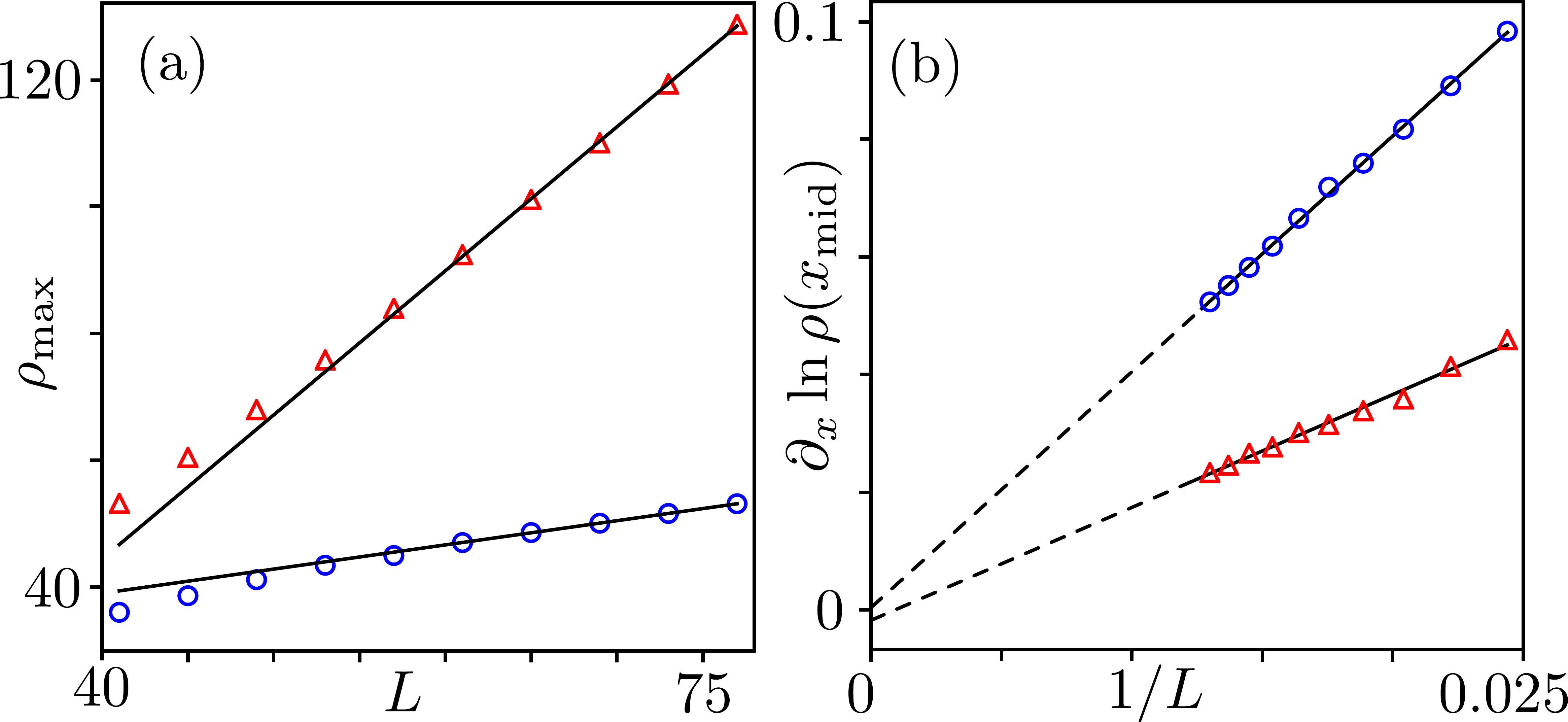}
\caption{
(a) Maximum density plotted as a function of system size $L$ for two values of $(\delta t_x/t_x, t_y/t_x)=(0.9, 0.04)$, shown using blue circles, and $(0.72, 0.72)$, shown as red triangles. For large systems, $\rho_{\rm max}\propto L$, as indicated by the black lines. (b) Slope of the density $\partial (\ln\rho_{\rm mid}) / \partial x$ for the same parameters. The slope is averaged over five neighboring lattice sites at the mid-point of the line connecting the dislocations. In the thermodynamic limit, the fit indicates a vanishing slope (within error bars), as shown by the black lines. \label{fig:scaling}}
\end{figure}

\emph{Scaling properties}.--- 
In order to show the extensive nature of the NHSE caused by the dislocation, we study numerically the length dependence of the maximum of the density in the vicinity of one dislocation.
For a 1D HN model, the maximum density is easily obtained from its conservation law $\int dr \rho_r = L$ which leads to $\rho_{\rm max}=(L/\xi)/(1-e^{-L/\xi})\approx L/\xi$. 
We find that also in the 2D case, the linear behavior of the maximum density persists [see Fig.~\ref{fig:scaling}(a)], and our results are consistent with $\rho_{\rm max}\propto L$ for a wide range of parameter values $t_y/t_x$ and $\delta t_x/t_x$. 
When keeping the system size constant, we observe that the maximum density is larger when $t_y\neq0$ compared to the decoupled limit.
This indicates that the dislocation-induced NHSE is different from that occurring in the 1D model, in the sense that states from the entire 2D system participate in the density collapse.

In contrast to the scaling behavior of the skin effect, the anti-skin effect saturates as a function of system size, with the minimum density $\rho_{\rm min}$ becoming independent of $L$ for large systems.
The precise values of $\rho_{\rm min}$ depend both on non-reciprocity as well as the strength of the vertical hoppings, but they are usually a few orders of magnitude smaller than the density far away from the dislocations, as can be seen also in Fig.~\ref{fig:WHN}(b).

Finally, we study the behavior of the density far from the cores of the dislocations.
We find that the density is roughly uniform as a function of the coordinates $x$ and $y$, with the largest variations occurring on a line connecting the two dislocations, corresponding to the 1D chain left over after removing $L$ sites from the system (the row of sites terminating at the dislocations in Fig.~\ref{fig:WHN}).
We examine the length dependence of the slope of the density at the midpoint of this chain, $\partial_x\ln\rho(x_{\rm mid})$.
When extrapolating to $L\to \infty$ for different $t_y/t_x$ and $\delta t_x/t_x$, we observe that the slope vanishes, indicating that in the thermodynamic limit the density profile is flat far away from the dislocation cores [see Fig.~\ref{fig:scaling}(b)].
This is in contrast to the density profile of the 1D Hatano-Nelson model, where $\rho_r \propto e^{r/\xi_{1D}}$ implies a nonzero $\partial_r \ln\rho(r_{\rm mid})  \sim 1/\xi_{1D}$, independent of system size.
The uniform density far from the dislocation cores is similar to the behavior of Hermitian systems hosting topological defects, where the defect is locally indistinguishable sufficiently far away from its core.
Here, however, ${\cal O}(L)$ states accumulate at one of the dislocations, and the remaining ${\cal O}(L^2 - L)$ appear to be distributed uniformly away from the defect cores.

\emph{Topological origin}.---
To determine the topological invariant associated with the dislocation-induced skin and anti-skin effects, we use the procedure outlined in Refs.~\cite{Sato2020, Kawabata2020}.
Starting from our NH system, we construct a doubled Hermitian Hamiltonian with chiral symmetry as
\begin{equation}\label{chiral-class}
 \widetilde{\cal H}=
 \begin{pmatrix} 
 0 & {\cal H}^{\phantom\dag} \\ 
 {\cal H}^\dag & 0 
 \end{pmatrix}.
\end{equation}
{As shown in Ref.~\cite{Sato2020}, the topological invariant responsible for the NHSE in ${\cal H}$ is identical to the one responsible for the existence of topologically protected mid-gap states in $\widetilde{\cal H}$.}

The doubled system is a stack of horizontal SSH chains \cite{Su1979}, coupled by diagonal hoppings. 
It belongs to the class BDI in the Altland-Zirnbauer classification \cite{Altland1997}, and has been discussed in Ref.~\cite{Hughes2020weakAIII}.
Using the Hermitian model Eq.~\eqref{chiral-class}, we find that the bulk of the system is gapped provided that $|t_y| < |t_x|$, and that each of the two dislocations traps an exponentially-localized zero-energy mode.
These mid-gap modes are protected by chiral symmetry, which forces the spectrum to be symmetric around $E=0$.
As a consequence, the energy of the dislocation-bound states cannot be shifted away from zero unless the bulk gap closes, or the two defects are brought in proximity of each other, such that their bound states can hybridize and gap out.

The doubled model provides a convenient starting point to demonstrate the topological origin of the NHSE caused by the dislocations.
In Hermitian systems, it has been shown that protected mid-gap states can appear not only at the boundaries but can be also trapped by topological defects.
In the latter case, their existence relies on the topological properties of defect Hamiltonians ${\cal H}({\bf k}, {\bf r})$, with $\bf{r}$ spanning a $D$-dimensional sphere around the defect, as proposed by Teo and Kane \cite{teo2010}. 
At each point $\bf{r}$ far away from the defect, ${\cal H}({\bf k}, {\bf r})$ is well-approximated by a Bloch Hamiltonian since the lattice is locally translation symmetric. 
However, upon encircling around the dislocation as indicated in Fig.~\ref{fig:WHN}(a), an extra translation by the Burgers vector ${\bf B}$ is needed as compared to a similar loop in the absence of the dislocation.

For our 2D model in the presence of a dislocation we can devise a NH defect Hamiltonian ${\cal H}({\bf k}, s)$ with 2D momenta ${\bf k}$ and a variable $s\in[0,1]$ which describes the loop around the topological defect.
The doubled Hamiltonian $\widetilde{\cal H}({\bf k}, s)$ corresponds to
a 2D chiral-symmetric system with a point defect, which is characterized by the 3D winding number, 
\begin{equation}
    W_3=\frac{ \epsilon^{\mu\nu\lambda} }{3} \int
    \frac{d^2{\bf k}ds  }{(2\pi i)^3}
   \: {\rm Tr}\big( q\partial_\mu q^\dag\: q\partial_\nu q^\dag 
    \: q\partial_\lambda q^\dag
   \big),
    \label{eq:3d-winding}
\end{equation}
according to the topological classification of Ref.~\cite{teo2010}.
The integration is over $T^2\times S^1$ (the effective 3D Brillouin zone of the defect Hamiltonian), $\epsilon^{\mu\nu\lambda}$ is the fully anti-symmetric Levi-Civita tensor, and $q({\bf k},s)={\cal H}/({\cal H}{\cal H}^{\dag})^{\!\text{\sfrac{1}{2}}}$ results after flattening the bands of the Hermitian Hamiltonian $\widetilde{\cal H}({\bf k}, s)$. 
Since our WHN model has only one band, $q({\bf k},s)$ is a scalar, which means that the winding number in Eq.~\eqref{eq:3d-winding} vanishes identically (due to the Levi-Civita tensor). 
This means that $W_3$ fails to capture the existence of topological zero modes of the doubled defect Hamiltonian $\widetilde{\cal H}$, and thus also fails to capture the dislocation-induced NHSE.

This motivates us to formulate an alternative approach, where instead of working with the defect Hamiltonian, we displace locally the Bloch wave functions as $u_{{\bf k},s}({\bf r}) = u^{0}_{{\bf k}}({\bf r}-s {\bf B})$ along the loop surrounding the dislocation, where $u^{0}_{{\bf k}}$ denotes the Bloch state of the defect-free crystal.
{That way the interplay of defected lattice represented by the Burgers vector and band structure topology are directly introduced through the displaced Bloch wave functions.} Consequently, the Berry connection of the system in the presence of a topological defect reads
\begin{equation}
    {\bf A}({\bf k},s)=
    \langle u_{{\bf k},s}|{\bm \nabla}_{{\bf k},s} | u_{{\bf k},s} \rangle
    ={\bf A}^0({\bf k}) + 
\hat{\bf z} A_s({\bf k}),
    \label{eq:berry-connection}
\end{equation}
{consisting of a defect-free part
${\bf A}^0({\bf k})=\langle u^0_{{\bf k},s}|\nabla_{\bf k} | u^0_{{\bf k},s} \rangle$
and an additional term due to the defects.
We find that the defect-free part is given by ${\bf A}^0({\bf k})=\frac{1}{2} q_{\bf k}^{-1} \nabla_{\bf k}q_{\bf k}$ with $q_{\bf k}=\varepsilon_{\bf k}/|\varepsilon_{\bf k}|$,
since the Bloch states for the doubled Hamiltonian are given by $|u^{0}_{{\bf k}}\rangle = (1/\sqrt{2})(q_{\bf k},1)^T$ .
The defect-enforced term $A_s({\bf k})=\langle u_{{\bf k},s}|\partial_s | u_{{\bf k},s} \rangle$
in Eq. \eqref{eq:berry-connection}, which originates from displacing Bloch states by the fictitious momentum $s$, can be written as} \cite{teo2010,SM}
\begin{equation}
A_s({\bf k})=-{\bf B}\cdot\langle u^{0}_{{\bf k}}| \nabla_{\bf r} |u^{0}_{{\bf k}}\rangle=i\,{\bf B}\cdot{\bf k}-{\bf B}\cdot{\bf a}^{p}({\bf k}),
\end{equation}
where ${\bf a}^{p}({\bf k})=\langle u^{0}_{{\bf k}}| (\nabla_{\bf r}+i{\bf k}) |u^{0}_{{\bf k}}\rangle$ is periodic over the reciprocal space \cite{blount1962}.
Now, associated with the Berry connection \eqref{eq:berry-connection}, there exist a Chern-Simons (CS) form ${\cal Q}_3 ({\bf k},s)=  d^2{\bf k}  ds \, \epsilon^{\mu\nu\lambda} \,{\rm Tr}\big( A_\mu \partial_\nu A_\lambda +\frac{2}{3} A_\mu A_\nu A_\lambda \big)$. 
From a detailed derivation \cite{SM}, we find a $\mathbb{Z}_2$, CS invariant,
\begin{eqnarray}
\vartheta=\frac{1}{(2\pi)^2}\, \int_{T^2\times S^1}   {\cal Q}_3 =\frac{1}{2}\hat{\bf z}\cdot\big({\bf B}\times {\bm \nu}\big) \!\!\!\! \mod 1,
\label{eq:theta-invariant}
\end{eqnarray}
determined by the cross product of the Burgers vector, and the 2D weak indices of WHN model, Eq.~\eqref{eq:average-windings}. The $\mathbb{Z}_2$ invariant $\vartheta\in\{ 0 , 1/2 \}$ can distinguish between the trivial and nontrivial phases of the doubled Hamiltonian in the presence of topological defects.~$\vartheta=1/2$ implies an odd parity of zero modes at a dislocation in the doubled Hamiltonian, and consequently the skin effect of topological defects in the WHN model.
The two dislocations shown in Fig.~\ref{fig:WHN} are characterized by Burgers vectors ${\bf B}=(B_x, B_y)=(0, \pm 1)$, which together with weak indices ${\bm \nu}=(1, 0)$ for $\delta t_x>0$, imply a half-integer $\vartheta$.
Remarkably, both the skin \emph{and} anti-skin effects are associated with a nontrivial invariant, signaling that also the localized density depletion around one of the dislocations is topological in origin.

{
The topological invariant Eq.~\eqref{eq:theta-invariant} is reminiscent of that characterizing dislocations in Hermitian weak topological phases \cite{teo2010}. The invariant there, also involves both the Burgers vector and the vector of weak indices, though their scalar
product instead of the cross product. As in that case, an intuitive understanding can be gained by going to the decoupled limit, where the Burgers vector specifies which type of termination is introduced into one of the chains, and the weak invariants control which terminations will show the NHSE. Furthermore, due to the correspondence, unraveled in Ref.~\cite{Sato2020}, between the NHSE and the topology of Hermitian systems given by 
Eq.~\eqref{chiral-class}, the invariant $\vartheta$ can be equally 
used for the 2D Hermitian model.
Namely, we can use it to predict the appearance of zero modes bounded to dislocations
in a 2D model consisting of arrays of SSH chains.
}

We emphasize that the CS invariants are not gauge-invariant in general, hence, $\vartheta$ only captures the integer/half-integer parity of the ${\mathbbm Z}/2$ invariant $\hat{\bf z}\cdot({\bf B}\times {\bm \nu})/2$.
As such, it can only distinguish the even or odd parity of dislocation-bound zero modes of the doubled, Hermitian model $\widetilde{\cal H}$, and not their total number.
This can be better understood by noting that for a single-band 2D NH model, the Berry connection ${\bf A}({\bf k},s)$ is a 3D Abelian gauge field and consequently, the CS invariant $\vartheta$ is identical to the \emph{Hopf index} \cite{Moore-hopf, deng2013-hopf, kennedy2016, liu-hopf}.
The Hopf index is generally a ${\mathbb Z}_{2N_{\rm Ch}}$ invariant where the integer $N_{\rm Ch}={\rm GCD}(C_x,C_y,C_s)$ is the greatest common divisor of three Chern numbers $C_j=\int d^2k_{\perp} {\Omega}_j$ defined over three 2D cross sections of $T^2\times S^1$. 
Using the definition of Berry curvature $\Omega_j=\varepsilon_{jkl}\partial_k A_l$ and the Berry connection given by Eq.~\eqref{eq:berry-connection}, we find $(C_x,C_y,C_s)=(B_y,-B_x,0)$ with ${\rm GCD}=1$, which justifies the ${\mathbb Z}_{2}$ nature of the topological invariant $\vartheta$ \cite{SM}. 
The correspondence between the WHN model with topological defects and Hopf insulators indicates that it also lies in the nonstable regime (the numbers of energy bands are below the bounds introduced in Ref. \cite{kennedy2016bott}).
As a result, it falls beyond the periodic table of stable phases where 2D chiral-symmetric systems with point defects are characterized with the winding number $W_3\in{\mathbb Z}$ \cite{teo2010}.

Finally, we find that, consistent with the topological invariant of Eq.~\eqref{eq:theta-invariant}, the density peak and dip shown in Fig.~\ref{fig:WHN} lose their robustness when ${\bf B}\times {\bm \nu}=0$, or when the point gap at $\varepsilon=0$ closes ($|t_y| \geq |t_x|$). We explore these cases in Supplemental Material \cite{SM}, comparing the nontrivial point defects studied above with trivial point defects, such as vacancies. We find that, while localized features in the density are still observed, for trivial point defects they are much weaker, and {can be suppressed by means of local perturbations.} They are similar to the conventional impurity states present in Hermitian systems, which do cause localized features in the local density of states, but do not show the robustness associated to the nontrivial topology.

\emph{Conclusions}.--- 
Using a prototypical 2D non-Hermitian system, we have shown that dislocations result in the formation of a non-Hermitian skin effect, signaled by the accumulation of density due to the localization of a macroscopically large number of states towards the dislocation.
This effect is observed numerically, and then understood using a topological invariant that falls outside the conventional, stable bulk-defect correspondence. {The analytical study of this density acummulation as well as its extensions to arrays of dislocations provides an interesting avenue for further research.}

Unlike the skin effect associated with the boundaries of non-Hermitian systems, we have found that dislocations can also host the anti-skin effect.
This is a topologically protected \emph{depletion} of an otherwise uniform density profile, which occurs at the dislocation core.
The possibility of an anti-skin effect offers an additional tool for tailoring the positions of eigenstates in non-Hermitian systems, and may be useful in the design of practical applications, such as sensors \cite{Budich2020} or light funnels \cite{Weidemann2020}.

\begin{acknowledgments}
\emph{Acknowledgments}.---
We thank C.~L.~Kane and A.~R.~Akhmerov for enlightening discussions, as well as Ulrike Nitzsche for technical assistance.
This work was supported by the Deutsche Forschungsgemeinschaft~(DFG, German
Research Foundation) under Germany's Excellence Strategy through the
W\"{u}rzburg-Dresden Cluster of Excellence on Complexity and Topology in Quantum
Matter -- \emph{ct.qmat} (EXC 2147, project-id 390858490), as well as through the DFG grant FU 1253/1-1. A.G.M. acknowledges financial support from Iran Science Elites
Federation under Grant No. 11/66332.
\end{acknowledgments}

\emph{Note added}.--- During the final stages of this work, a related paper, Ref.~\cite{Schindler2021}
came out in which the authors observe the presence of skin effects at dislocations in NH systems, including for a model that is equivalent to our WHN system.

\bibliography{topo_defect}
\end{document}


\title{Supplemental Material:\\ Non-Hermitian skin effect of dislocations and its topological origin}

\author{Balaganchi A. Bhargava}
\affiliation{Institute for Theoretical Solid State Physics, IFW Dresden and W\"urzburg-Dresden Cluster of Excellence ct.qmat, Helmholtzstr. 20, 01069 Dresden, Germany}

\author{Ion Cosma Fulga}
\affiliation{Institute for Theoretical Solid State Physics, IFW Dresden and W\"urzburg-Dresden Cluster of Excellence ct.qmat, Helmholtzstr. 20, 01069 Dresden, Germany}

\author{Jeroen van den Brink}
\affiliation{Institute for Theoretical Solid State Physics, IFW Dresden and W\"urzburg-Dresden Cluster of Excellence ct.qmat, Helmholtzstr. 20, 01069 Dresden, Germany}
\affiliation{Institute  for  Theoretical  Physics,  TU  Dresden,  01069  Dresden,  Germany}

\author{Ali G. Moghaddam}
\affiliation{Department of Physics, Institute for Advanced Studies in Basic Sciences (IASBS), Zanjan 45137-66731, Iran}
\affiliation{Computational Physics Laboratory, Physics Unit, Faculty of Engineering and
Natural Sciences, Tampere University, P.O. Box 692, FI-33014 Tampere, Finland}
\affiliation{Institute for Theoretical Solid State Physics, IFW Dresden and W\"urzburg-Dresden Cluster of Excellence ct.qmat, Helmholtzstr. 20, 01069 Dresden, Germany}

\begin{abstract}
    In this Supplemental Material, we first provide detail derivation of the Chern-Simons topological invariant. Specially we show how it reduces to the the cross product of the dislocation Burgers vector and the weak indices of the 2D WHN model. Next, we consider the role of non-topological defects for which show the localized features in the density are drastically weaker and fragile in the presence of local perturbations.
\end{abstract}

\date{\today}

\maketitle

\section{Topological invariant}
\label{app:top_inv}

As discussed in the main text, we can double the single-band NH model, which yields a two-band Hamiltonian with chiral symmetry
\begin{eqnarray}
\widetilde{\cal H}_{\bf k}=\begin{pmatrix}
0 & \varepsilon_{\bf k}  \\ \varepsilon_{\bf k}^\ast  &0  
\end{pmatrix}.\label{eq:H-chiral}
\end{eqnarray}
with
$\varepsilon_{\bf k} = 2 \,\big( t_x\,\cos k_x + t_y\,\cos k_y+i \,\delta t_x \,\sin k_x\big)$
for the 2D WHN model. 
In the defect-free case, the Bloch functions of the Hamiltonian Eq.~\eqref{eq:H-chiral} read 
\begin{eqnarray}
|u^{0}_{{\bf k}}\rangle = \frac{1}{\sqrt{2}}\begin{pmatrix}
e^{i\varphi_{\bf k}}  \\ 1
\end{pmatrix},\label{eq:pure-bloch}
\end{eqnarray}
with $e^{i\varphi_{\bf k}}=q_{\bf k}=\varepsilon_{\bf k}/|\varepsilon_{\bf k}|$, such that the Berry connection is
\begin{align}
    {\bf A}^0({\bf k})&=
    \frac{1}{2} q_{\bf k}^{-1} \nabla_{\bf k}q_{\bf k} =  \frac{i}{2} \nabla_{\bf k} \varphi_{\bf k}.
\end{align}

\subsection{Dislocation-induced terms}

Next, we study the effect of the dislocation by displacing the Bloch states as $u_{{\bf k},s}({\bf r}) = u^{0}_{{\bf k}}({\bf r}-s {\bf B})$. That way, it is clear that we have an effective 3D Brillouin zone (BZ) $T^2\times S^1$
-- consisting of two momenta $k_x$, $k_y$, and a third fictitious momentum $s$ --
for the 2D system with a point defect.
The Berry connection associated with this 3D BZ will have an extra term due to the $s$-dependence of the displaced Bloch eigenstates, which gives  
\begin{align}
A_s({\bf k})=\langle u_{{\bf k},s}| \partial_{s} |u_{{\bf k},s}\rangle =
-{\bf B}\cdot\langle u^{0}_{{\bf k}}| \nabla_{\bf r} |u^{0}_{{\bf k}}\rangle, \label{eq:Berry-A_s}
\end{align}
using the chain rule $\partial_s u({\bf r}-s{\bf B})=-{\bf B}\cdot\nabla_{\bf r} u$.
The matrix elements of the operator $\nabla_{\bf r}$ with respect to the Bloch states decompose into two parts.
The first term is nothing but the corresponding matrix element of $i{\bf k}$ (which is identical with $\nabla_{\bf r}$ in vacuum). But there will be a second term ${\bf a}^p({\bf k})$ that can be easily seen by expanding Bloch states in terms of their Fourier components as 
\begin{equation}
\langle{\bf r}|u^0_{n{\bf k}}\rangle=\sum_{\bf G} e^{i{\bf G}\cdot {\bf r}} \: {\cal U}^{(n)}_{{\bf k+G}} ,    
\end{equation}
for a general multi-band system ($n$ denotes the $n^{\rm th}$ occupied band). 
For the matrix elements of $\nabla_{\bf r}$, this results in
\begin{eqnarray}
\langle u^0_{m{\bf k}} | \nabla_{\bf r} |u^0_{n{\bf k}} \rangle &=&
\sum_{\bf G} i\,{\bf G}\: 
{\cal U}^{(m)\,\ast}_{{\bf k+G}}\:
{\cal U}^{(n)}_{{\bf k+G}}
~\nonumber\\
&=&-i\, {\bf k}\, \delta_{mn} + {\bf a}^p_{mn}({\bf k}),\label{eq:periodic-a^p}
\end{eqnarray}
with 
\begin{equation}
{\bf a}^p_{mn}({\bf k})=
i\sum_{\bf G} ({\bf k+G})\: 
{\cal U}^{(m)\,\ast}_{{\bf k+G}}\:
{\cal U}^{(n)}_{{\bf k+G}} ~,
\end{equation}
which is periodic under translating by any reciprocal lattice vector ${\bf G}$.
The above identity, which was first noticed by Blount in the 60's, is a bit less familiar than the corresponding relation for the position operator ${\bf r}=i{\bf k}+{\bf A}^0_{mn}({\bf k})$ with non-Abelian Berry connection matrix elements ${\bf A}^0_{mn}({\bf k})$.
Plugging the identity Eq.~\eqref{eq:periodic-a^p} into Eq.~\eqref{eq:Berry-A_s}, we find
\begin{align}
A_s({\bf k})=i\,{\bf B}\cdot{\bf k}-{\bf B}\cdot{\bf a}^{p}({\bf k}),
\end{align}
which is identical to the Eq.~(7) presented in the main text.
Note that our Hamiltonian Eq.~\eqref{eq:H-chiral} has only a single occupied band, and therefore we can simply drop the band indices. 

\subsection{Chern-Simons form and topological invariant}

We now turn to the Chern-Simons (CS) 3-form ${\cal Q}_3 ({\bf k},s)=  d^2{\bf k}  ds \, \epsilon^{\mu\nu\lambda} \,{\rm Tr}\big( A_\mu \partial_\nu A_\lambda +\frac{2}{3} A_\mu A_\nu A_\lambda \big)$ associated with the Berry connection ${\bf A}({\bf k},s)= {\bf A}^0({\bf k})+\hat{\bf z}\,A_s({\bf k})$. 
Since we have a single occupied band, the corresponding Berry connection is Abelian and the CS form simplifies to
\begin{align}
    {\cal Q}_3 ({\bf k},s) &=  d^2{\bf k}  ds \: \epsilon^{\mu\nu\lambda}  \: A_\mu \partial_\nu A_\lambda \nonumber \\ &= d^2{\bf k}  ds \: {\bf A}\cdot \big ({\bm \nabla}_{{\bf k},s}\times{\bf A} \big) 
\end{align}
Note we use a convention such that ${\bm \nabla}_{{\bf k},s}\equiv (\partial_{k_x},\partial_{k_y},\partial_{s})$.
The term coming from the Bloch bands ${\bf A}^0({\bf k})$ is a gradient field, and therefore curl-free. 
So we only need to consider the curl of the last term $\hat{\bf z}\,A_s({\bf k})$ which reads
\begin{align}
{\bm \nabla}\times{\bf A}&=
-i \hat{\bf z} \times {\bf B}+\big( \hat{\bf z} \times \nabla_{\bf k} \big) {\bf B}\cdot{\bf a}^p
,
\end{align}
and consequently we obtain
\begin{align}
{\bf A} \cdot \big({\bm \nabla}_{{\bf k},s}\times{\bf A}\big)
= i \hat{\bf z} \cdot \big[{\bf A}^0({\bf k})\times {\bf B}\big] - {\bf Q}^p({\bf k})\cdot{\bf B}. \label{eq:final-CS-form}
\end{align}
The second term containing
${\bf Q}^p({\bf k})=  \hat{\bf z} \cdot \big(
{\bf A}^0\times \nabla_{\bf k}
\big) {\bf a}^p({\bf k})$ is, in fact, a periodic function of 2D momenta ${\bf k}$ and, therefore, its integral
over the full 3D BZ $T^2\times S^1$ identically vanishes. So we obtain
\begin{align}
    \vartheta&=\frac{1}{(2\pi)^2}\, \int_{T^2\times S^1}   {\cal Q}_3 
    \nonumber\\
    &=
    \hat{\bf z}\cdot \bigg[ {\bf B} \times \int_0^1 ds  \int \frac{d^2{\bf k} }{i(2\pi)^2} 
\: {\bf A}^0({\bf k}) \bigg]
 \nonumber\\
    &=
    \frac{1}{2}\hat{\bf z}\cdot \bigg[ {\bf B} \times   \int \frac{d^2{\bf k} }{i(2\pi)^2} 
\: q_{\bf k}^{-1} \nabla_{\bf k} q_{\bf k} \bigg],
\end{align}
which reduces to the Eq.~(8) in the main text, noting that 
\begin{equation}
\int \frac{d^2{\bf k} }{i(2\pi)^2} 
\: q_{\bf k}^{-1} \nabla_{\bf k} q_{\bf k}\equiv
\int \frac{d^2{\bf k} }{i(2\pi)^2} 
\: \varepsilon_{\bf k}^{-1 }\nabla_{\bf k}\varepsilon_{\bf k}={\bm \nu}.
\end{equation}

\subsection{Role of gauge transformations}

Here we consider the role of choosing a different gauge in writing the Berry connection. 
In general, the Berry connection and its curvature under a gauge transformation can be written as
\begin{eqnarray}
&&{\bf A}^\prime=g^{-1}{\bf A} g+g^{-1} {\bm \nabla} g, \\
&&{\bm \Omega}^\prime=g^{-1}{\bm \Omega} g  .
\end{eqnarray}
In the simplest case of an Abelian structure for the Berry connection, we deal with $U(1)$ gauges of the form $g=e^{i\chi}$ with a general phase function $\chi({\bf k},s)$ defined over the $T^2 \times S^1$ BZ. 
Consequently the gauge transformation becomes 
\begin{eqnarray}
&&{\bf A}^\prime= {\bf A} + i \, {\bm \nabla} \chi \\
&&{\bm \Omega}^\prime= {\bm \Omega}   
\end{eqnarray}
Then, the CS form for the transformed Berry connection ${\bf A}^{\prime}$ reads
\begin{align}
{\cal Q}^\prime_3 
&= {\cal Q}_3 + i\, d^2{\bf k}ds\:{\bm \nabla}\chi \cdot \big( {\bm \nabla} \times {\bf A}\big).
\end{align}
The new term caused by the gauge transformation
can be decomposed as
\begin{align}
i{\bm \nabla}\chi \cdot\big( {\bm \nabla} \times {\bf A}\big)
=i \hat{\bf z}\cdot  \big( {\nabla}_{\bf k}\chi \times {\bf B} \big)- \hat{\bf z}\cdot 
\big( {\bm \nabla}  \chi \times {\bm \nabla} \big)   {\bf B}\cdot {\bf a}^p,
\end{align}
where the latter part is periodic, and its integral vanishes. Therefore, we obtain  
\begin{align}
\frac{1}{(2\pi)^2}\int \big( {\cal Q}^\prime_3 -{\cal Q}_3 \big)  = \hat{\bf z}\cdot 
\big( {\bf B} \times {\bm w}_{\chi}
 \big) \label{eq:gauge-winding}
\end{align}
with averaged 1D winding number
\begin{equation}
{\bm w}_{\chi}=\int_0^1 ds  \int \frac{d^2{\bf k} }{i(2\pi)^2}     {\nabla}_{\bf k}\chi ,
\end{equation}
of the gauge variable $\chi({\bf k},s)$.
The additional term due to the gauge transformation given by Eq.~\eqref{eq:gauge-winding} is integer-valued, which justifies the gauge-invariance of $\vartheta$
defined as a ${\mathbb Z}_2$ topological invariant.

\section{Non-topological defects}
\label{app:defects}

\begin{figure}[h!]
\includegraphics[width=0.45\linewidth]{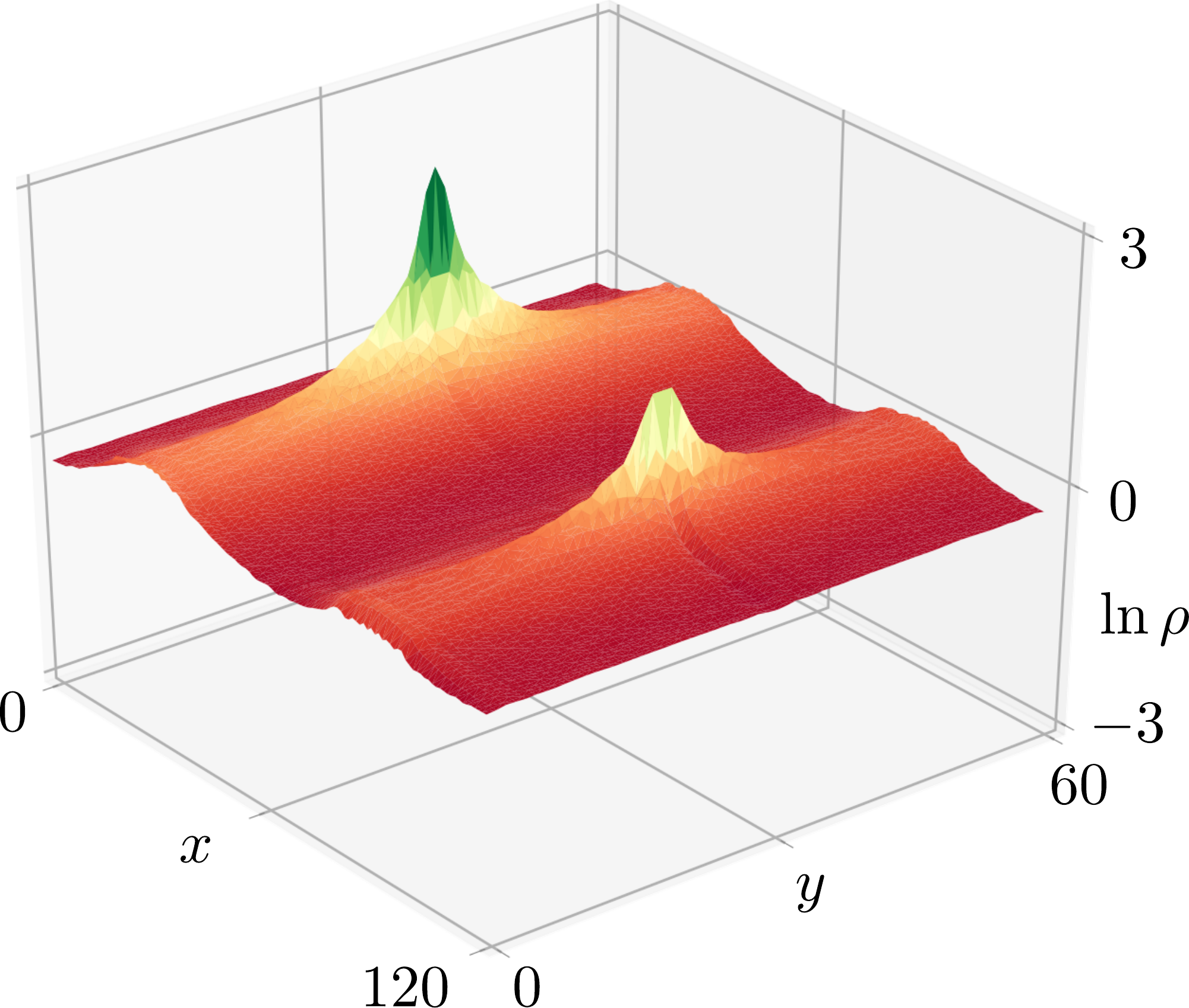} 
\caption{
Plot of the density using $\delta t_x/t_x=0.6$ and $t_y/t_x=3$. \label{fig:gapless}}
\end{figure}

Here we comment on the behavior of defects which are not characterized by a nontrivial invariant $\vartheta$. We focus on the case of a gapless system, $|t_y| > |t_x|$, on a gapped system for which ${\bf B}\times {\bm \nu}=0$, as well as on a single vacancy.
If the system is gapless, we find that the anti-skin effect disappears. There are now two pronounced peaks in the density, as shown in Fig.~\ref{fig:gapless} for $\delta t_x/t_x=0.6$ and $t_y/t_x=3$. Further, there is significant variation in the density away from the defect cores, which extends through the entire system in the $y$ direction. Note that the linear dimension of the system considered here is twice larger than the one shown in Fig.~1 of the main text.

Next, we consider a gapped system, $|t_y| < |t_x|$, but which hosts dislocations obeying ${\bf B}\times {\bm \nu}=0$. 
This is achieved using the same cut and glue procedure outlined in the main text. 
Starting from a rectangular system consisting of $L \times 2L$ sites in the $x$ and $y$ directions (opposite aspect ratio as compared to the main text), we remove a column of $L$ sites, and then glue the resulting cut back together using the horizontal, non-reciprocal hoppings.
Also in this case, we observe that the anti-skin effect disappears, and two symmetric density peaks are localized at the two dislocation cores.
These peaks, however, are sensitive to the Hamiltonian parameters, as can be intuitively seen by considering the decoupled limit, $t_y=0$.
In this limit, density peaks vanish despite of the non-reciprocal hoppings, since each of the decoupled Hatano-Nelson models retains its periodic boundary conditions, such that $\rho_{\bf r}=1$ due to the translation symmetry of each individual chain.

\begin{figure}[t]
\includegraphics[width=0.45\linewidth]{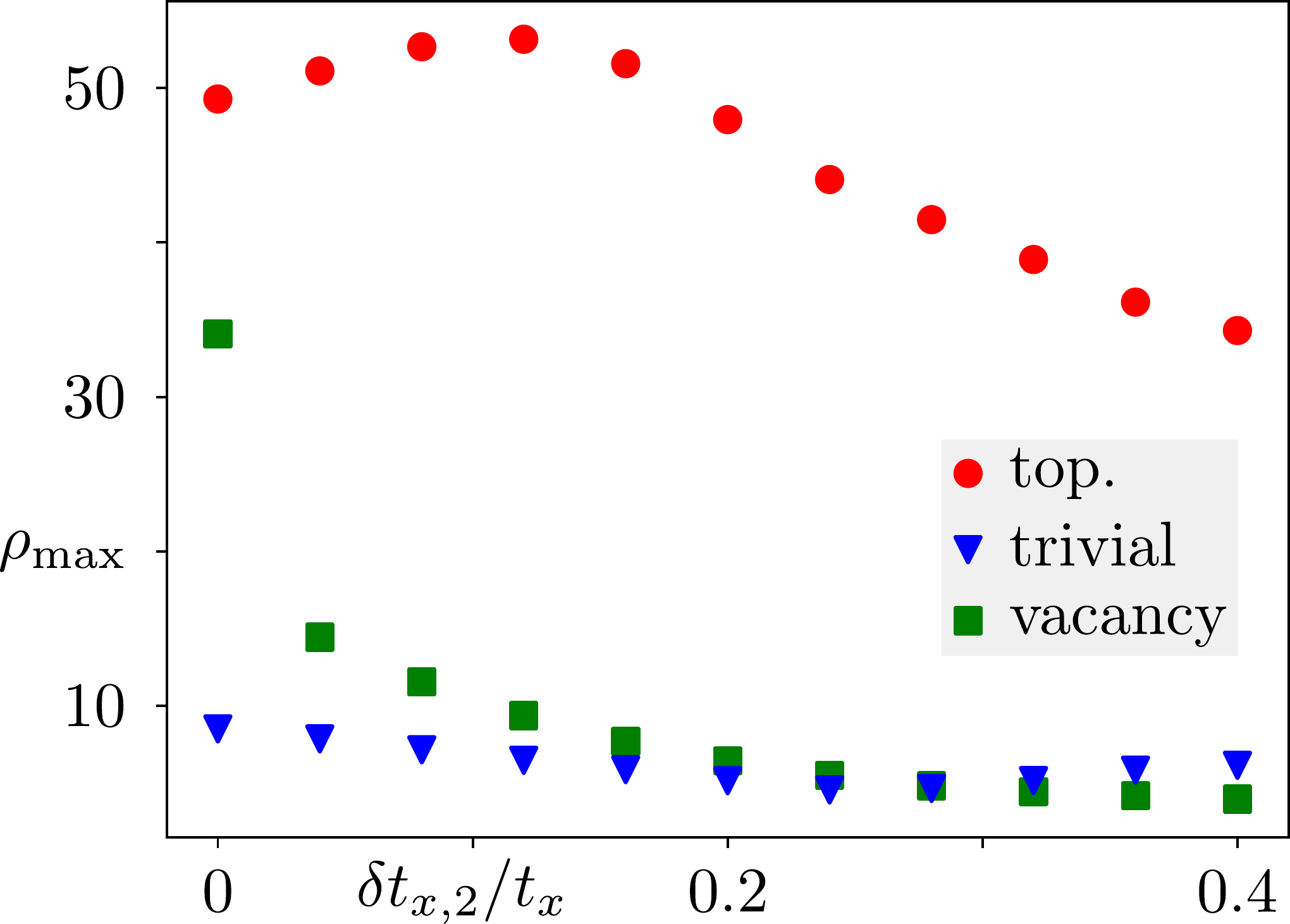} 
\caption{
Dependence of the maximum density $\rho_{\rm max}$ on the strength of a next-nearest-neighbor, horizontal non-reciprocal hopping, $t_{x,2}$. Red circles denote the behavior of nontrivial dislocations, for which ${\bf B}\times {\bm \nu}\neq0$, blue triangles show trivial dislocations, ${\bf B}\times {\bm \nu}=0$, and green squares correspond to a vacancy.
\label{fig:t2}}
\end{figure}

Finally, we examine the behavior of a vacancy, a nontopological defect introduced by removing a single site from the system. 
In this case, the density shows an adjacent peak and dip, but these features are highly sensitive to changing Hamiltonian parameters.
We show this in Fig.~\ref{fig:t2}, where we look at the maximum density upon adding a next-nearest-neighbor non-reciprocal hopping in the $x$ direction, $t_{x,2}$.
Upon increasing $t_{x,2}$, the maximum density at the vacancy shows a pronounced decrease.
This is in contrast to the behavior of the topological skin effect present at dislocations for which ${\bf B}\times {\bm \nu}\neq0$, which increases as a function of the next-nearest-neighbor hopping, at least when $\delta t_{x,2}$ is small.

\bibliography{topo_defect}